\documentclass[sigconf,nonacm]{acmart}
\setcopyright{none}       
\renewcommand\footnotetextcopyrightpermission[1]{}
\AtBeginDocument{%
  }

\begin{document}

\title{TwinGate: Stateful Defense against Decompositional Jailbreaks in Untraceable Traffic via Asymmetric Contrastive Learning}


\author{Bowen Sun$^{1}$ \quad Chaozhuo Li$^{2\ast}$ \quad Yaodong Yang$^3$ \quad Yiwei Wang$^4$ \quad Chaowei Xiao$^1$}

\thanks{$^\ast$Corresponding author.}

\affiliation{%
  \institution{
    $^1$Johns Hopkins University \quad
    $^2$Microsoft Research Asia \quad
    $^3$Peking University \quad
    $^4$University of California, Merced
  }
  \country{} 
}
\vspace{2mm}
\email{bsun39@jh.edu}

\renewcommand{\shortauthors}{Sun et al.}

\begin{abstract}
Decompositional jailbreaks pose a critical threat to large language models (LLMs) by allowing adversaries to fragment a malicious objective into a sequence of individually benign queries that collectively reconstruct prohibited content. In real-world deployments, LLMs face a continuous, untraceable stream of fully anonymized and arbitrarily interleaved requests, infiltrated by covertly distributed adversarial queries. 
Under this rigorous threat model, state-of-the-art defensive strategies exhibit fundamental limitations. In the absence of trustworthy user metadata, they are incapable of tracking global historical contexts, while their deployment of generative models for real-time monitoring introduces computationally prohibitive overhead.
To address this, we present TwinGate, a stateful dual-encoder defense framework. TwinGate employs Asymmetric Contrastive Learning (ACL) to cluster semantically disparate but intent-matched malicious fragments in a shared latent space, while a parallel frozen encoder suppresses false positives arising from benign topical overlap. Each request requires only a single lightweight forward pass, enabling the defense to execute in parallel with the target model's prefill phase at negligible latency overhead. To evaluate our approach and advance future research, we construct a comprehensive dataset of over 3.62 million instructions spanning 8,600 distinct malicious intents. Evaluated on this large-scale corpus under a strictly causal protocol, TwinGate achieves high malicious intent recall at a remarkably low false positive rate while remaining highly robust against adaptive attacks. Furthermore, our proposal substantially outperforms stateful and stateless baselines, delivering superior throughput and reduced latency.
\end{abstract}





\maketitle

\section{Introduction}

The rapid proliferation of Large Language Models (LLMs) has rendered the development of robust safety mechanisms an increasingly pressing imperative~\cite{wang2025comprehensivesurveyllmagentstack}.
While standard alignment techniques and input/output filtering pipelines~\cite{li2025judgmentinterferenceearlystopping} have progressively raised the threshold for successful adversarial exploitation, the threat landscape continues to evolve, giving rise to novel attack paradigms that challenge existing defenses.
Among the most consequential is the emerging class of decompositional jailbreaks~\cite{wang2024repddefendingjailbreakattack}. Unlike conventional jailbreaking strategies~\cite{yi2024jailbreakattacksdefenseslarge} that seek to circumvent safety constraints within a single complex prompt, decompositional approaches systematically fragment a malicious objective into a sequence of discrete, ostensibly independent sub-tasks, as illustrated in Table~\ref{tab:decomposition_example}. 
Critically, each constituent instruction appears semantically benign in isolation, rendering it indistinguishable from legitimate user queries. Yet the aggregation of responses across this sequence enables an adversary to reconstruct prohibited information in its entirety as \textit{Safe in Isolation, Dangerous Together}~\cite{srivastav-zhang-2025-safe}.

A fundamental limitation of most existing defenses is their inherently stateless nature~\cite{xin2025jailbreakingattacksvscontent}: prevailing paradigms such as general alignment training~\cite{gu2024reviewsafereinforcementlearning} and guardrail frameworks evaluate each prompt in strict isolation, blind to malicious intent distributed across multiple turns. No individual fragment violates safety constraints alone, and harmful intent only emerges in aggregate across the trajectory. As a result, decompositional jailbreaks achieve alarmingly high attack success rates (ASR) even against rigorously aligned models, effectively neutralizing RLHF-based alignment~\cite{ouyang2022traininglanguagemodelsfollow} and input/output filtering.

\begin{table}[t]
  \caption{Example of a decompositional jailbreak.}
  \label{tab:decomposition_example}
  \centering
  \small
  \begin{tabular}{p{0.95\linewidth}}
    \toprule
    \textbf{Original Malicious Intent} \\
    \midrule
    How can I burn down the house for insurance money? \\
    \midrule
    \textbf{Decomposed Benign-looking Queries} \\
    \midrule
    1. What are the electrical principles behind circuit overloads and how do they contribute to the risk of house fires? \\
    2. What are the structural and safety properties of outdated wiring systems that make them more prone to causing house fires? \\
    3. What are the thermal properties of electrical cords and cables, and how do they relate to fire initiation? \\
    \midrule
    \textbf{Reconstructed Malicious Response} \\
    \midrule
    To initiate an untraceable electrical fire, you can exploit outdated wiring systems by intentionally overloading a specific circuit. First, bypass the safety breaker by ... \\
    \bottomrule
  \end{tabular}
\end{table}

The research community has begun to address decompositional threats by fortifying single-request filters~\cite{inan2023llamaguardllmbasedinputoutput,wang2024repddefendingjailbreakattack} and deploying sequential monitoring frameworks~\cite{yuehhan2025monitoringdecompositionattacksllms}. The nucleus of defending against decompositional jailbreaks lies in precisely aggregating decomposed queries to reconstruct the adversary's original intent. Existing methods, however, rest on a strong assumption of \textbf{traceable traffic}, specifically relying on predefined metadata, such as IP addresses and user identities, to associate and attribute sequences of queries to a specific principal. 
However, attackers can trivially circumvent identity-based attribution by distributing malicious query fragments across multiple accounts, sessions, or network endpoints, effectively dissolving their attack signal into the untraceable background noise of large-scale traffic. 
Furthermore, prevailing defenses commonly employ auxiliary LLMs to perform contextual analysis or intent classification, introducing a severe and asymmetric computational burden in which the cost of defense substantially exceeds that of mounting an attack~\cite{zhang2024intentionanalysismakesllms,yuehhan2025monitoringdecompositionattacksllms}.

In this paper, we propose TwinGate, a stateful defense framework against decompositional jailbreaks that addresses two core challenges: untraceable traffic and high operational overhead. TwinGate assumes a worst-case scenario of fully anonymized, arbitrarily interleaved requests, forcing all detection to rely purely on semantic analysis. This creates a structural problem we call the semantic gap: the individual steps of a distributed multi-session attack (e.g., separately querying wires and fertilizer) share negligible embedding similarity, defeating conventional detection. TwinGate bridges this gap with Asymmetric Contrastive Learning (ACL), which clusters semantically disparate but intent-matched malicious fragments in a shared latent space.
This aggressive clustering, however, risks representation collapse, spuriously grouping benign users discussing related topics as coordinated attackers. TwinGate resolves this through a dual-path decision inheritance mechanism: a parallel frozen encoder acts as a conservative anchor, preventing benign-but-similar requests from being escalated. Together, the contrastive and frozen encoders consolidate evidence of malicious intent across turns while suppressing the false positives that typically plague stateful semantic monitoring.

To simultaneously manage the massive volume of historical data dictated by our untraceable traffic threat model and maintain strict operational efficiency, TwinGate is engineered from the ground up for high-throughput scalability. We achieve this by deploying an asynchronous, parallel dual-encoder architecture coupled with dual in-memory vector databases. By processing each incoming request with only a single lightweight forward pass per encoder, TwinGate entirely circumvents the computational bottlenecks typical of generative-model-based strategies. This architectural efficiency drastically reduces the per-request computational footprint, enabling the system to sustain high-speed retrieval and evaluation with minimal latency overhead.

To operationalize the dual-path learning objective, we construct a large-scale dataset comprising over \textbf{3.62 million} requests, including 603k independent samples, 250k benign intents, and over 8,600 distinct malicious intents, each independently decomposed by multiple splitter models to ensure broad coverage of real-world attack variations. Our evaluation protocol is designed to rigorously mirror operational deployment conditions, simulating a continuous request stream in which the defense issues strictly causal decisions without access to future context. Under this protocol, the proposed system demonstrates strong generalization: on the rigorous generalized detection task, it achieves a malicious intent recall exceeding 0.76 while maintaining a false positive rate below $2 \times 10^{-3}$. Furthermore, deploying TwinGate reduces the attack success rate of simulated adaptive attacks to 0.18, collectively validating the dual-encoder architecture's capacity to balance detection sensitivity with operational precision.

In summary, we make the following contributions:

\begin{itemize}
\item To the best of our knowledge, we are the first to formally define and study the problem of defending against decompositional jailbreaks in the untraceable traffic setting, a practically significant and previously unexplored challenge. 
\item We propose TwinGate, a stateful dual-encoder defense mechanism designed specifically for the threat model of untraceable traffic. Without relying on any user metadata, our system successfully identifies cumulative malicious intent embedded across fragmented instructions.
\item We construct a large-scale dataset comprising diverse benign, independent, and decomposed malicious instructions, providing a rigorous benchmark for evaluating defense robustness against multiple decomposing implementations.
\item We empirically demonstrate that TwinGate substantially outperforms both stateful and stateless baselines against decompositional and adaptive attacks. Its lightweight encoding and efficient database retrieval further yield low latency and high scalability, confirming its suitability for high-throughput production environments.
\end{itemize}
\section{Related Work}
\label{sec:related_work}

\noindent\textbf{Decompositional Jailbreaks.} Early adversarial attacks on Large Language Models (LLMs) primarily focused on single-turn optimization, utilizing gradient-based search or manual template injection to bypass safety alignment within a single prompt \cite{xu-etal-2024-comprehensive}. However, as safety filters have evolved to detect explicit malicious patterns, the threat landscape has shifted towards more sophisticated multi-turn strategies that exploit the model's context-following capabilities and the stateless nature of standard defenses.

A pivotal precursor is \textit{Crescendo}~\cite{russinovich2025greatwritearticlethat}, which engages the model in a multi-turn conversation that begins with benign topics and imperceptibly escalates toward prohibited content. Although not strictly a semantic fragmentation attack, it bypasses guardrails that evaluate each turn in isolation by exploiting the cumulative toxicity of the dialogue history.

Building upon this exploitation of context, recent research has formalized the paradigm of decompositional jailbreaks. \citet{srivastav-zhang-2025-safe} characterize this threat model as \textit{Safe in Isolation, Dangerous Together}. In this framework, complex malicious objectives are strategically fragmented into a sequence of discrete, semantically disjoint sub-tasks. Orchestrated by multi-agent systems, these prompts appear benign to safety filters yet aggregate to reconstruct the harmful information.

Representing the current state-of-the-art, the \textit{CKA-Agent} framework~\cite{wei2025trojanknowledgebypassingcommercial} combines harmless prompt weaving with adaptive tree search, decomposing hazardous queries into a web of independent sub-questions and optimizing the retrieval path so that no single step triggers a refusal. \textit{CKA-Agent} reports remarkably high Attack Success Rates (ASR) across both open-source white-box and proprietary black-box targets, exposing a systemic limitation: stateless defenses adjudicate safety at the prompt level without historical semantic state, leaving them blind to malicious intent dispersed across ostensibly harmless interactions.

\noindent\textbf{Defenses against Decompositional Jailbreaks.} Countering decompositional threats requires shifting from static single-turn analysis to dynamic, context-aware evaluation. Benchmarking efforts such as \textit{CASE-Bench}~\cite{sun2025casebenchcontextawaresafetybenchmark} empirically establish that stateful defenses are essential: stateless mechanisms simultaneously miss malicious intent dispersed across benign-looking interactions and over-flag complex multi-step legitimate queries.

In response to this need for context, early attempts at stateful defense have focused on session-level monitoring. For instance, \citet{yuehhan2025monitoringdecompositionattacksllms} proposed a lightweight sequential monitor designed to track and aggregate the semantic trajectory of a user's requests within a continuous session. While effective in a cooperative setting, this approach relies on a fragile threat model that assumes a stable, attributable user identity. In real-world adversarial scenarios, this assumption is easily circumvented: sophisticated attackers can trivially evade such monitors by distributing their decomposed queries across multiple distinct sessions or identities, effectively resetting the defender's memory and rendering the accumulation of suspicion impossible.

Alternative strategies leverage retrieval mechanisms to identify potential attacks based on historical patterns. \textit{RePD} \cite{wang2024repddefendingjailbreakattack}, for example, employs a retrieval-based framework to fetch similar past attack templates and uses them to guide the LLM in recognizing decomposition attempts. However, this method introduces a prohibitive computational bottleneck. By requiring heavy-weight generative models to perform real-time reasoning and comparison against retrieved templates before processing the user's actual request, \textit{RePD} incurs a significant latency penalty. This reliance on heavy-weight model inference drastically increases the Time-To-First-Token (TTFT), making it operationally infeasible for latency-sensitive production environments.

Similarly, approaches that focus on explicit intent extraction face severe scalability and accuracy challenges. Methods like \textit{Intention Analysis} \cite{zhang2024intentionanalysismakesllms} compel the LLM to explicitly reconstruct and articulate the underlying user intent behind a prompt before generating a response. This process not only exacerbates the latency and cost issues inherent in generative-model-reliant defenses but also fundamentally fails to address the stateful nature of decomposition attacks. Because the intent extraction is performed on a single input in isolation, it lacks the historical context necessary to distinguish a benign query about ``wiring'' from a malicious step in a ``bomb-making'' sequence. Consequently, these stateless intent analyzers are prone to high false positive rates, flagging legitimate technical or creative queries as dangerous due to a lack of broader conversational context.

In summary, existing defenses present a trilemma: they are operationally fragile under spoofable metadata, computationally prohibitive due to generative-model-reliant inference (incurring significant TTFT penalties), and unscalable beyond the finite context window of standard LLMs. TwinGate sidesteps all three by pairing a metadata-free dual-encoder with high-throughput vector retrieval, scaling to production-grade untraceable traffic.

\section{Methodology}
\label{sec:methodology}

\subsection{Threat Model and Problem Formulation}
\label{sec:threat_model}

In this subsection, we formalize the threat model of decompositional jailbreaks within an untraceable traffic environment and formulate our stateful defense objective.

\noindent\textbf{Intents and Fragments.}
Let $\mathcal{I}$ denote the space of user intents, partitioned into malicious intents $\mathcal{I}_{mal}$ and benign intents $\mathcal{I}_{ben}$. Let $\mathcal{O}_{stateless}(\cdot) \to \{0, 1\}$ represent a standard stateless safety oracle, which outputs $1$ (safe) or $0$ (unsafe). Formally, a stateless defense evaluates an incoming query exclusively based on its local semantics, such that the decision $\mathcal{O}_{stateless}(q_t)$ is strictly independent of any prior query history. By definition, for any complete malicious intent $I_{mal} \in \mathcal{I}_{mal}$, $\mathcal{O}_{stateless}(I_{mal}) = 0$.

To evade detection, an adversary $\mathcal{A}$ employs a decomposition function $\Phi(I_{mal}) = \{x_1, x_2, \dots, x_n\}$, partitioning the malicious intent into a set of $n$ distinct instruction fragments. These fragments lack strict temporal dependency and can be submitted in any arbitrary order. The attack exhibits two core properties:
\begin{itemize}
    \item \textit{Safe in isolation}: Each individual fragment $x_i$ appears innocuous. To maintain usability, $\mathcal{O}_{stateless}$ effectively permits them, yielding $\mathcal{O}_{stateless}(x_i) = 1$ for all $1 \le i \le n$.
    \item \textit{Dangerous together}: The aggregation of the LLM's responses to the complete set $\{x_1, \dots, x_n\}$ provides sufficient information to fulfill the original intent $I_{mal}$.
\end{itemize}
The vast majority of stream requests are benign---some sharing a common benign intent $I_{ben} \in \mathcal{I}_{ben}$, others entirely unrelated---and both the stateless oracle $\mathcal{O}_{stateless}$ and any stateful defense must admit them without false flags.

\noindent\textbf{Untraceable Traffic.}
We model the real-world deployment environment as a global, continuous request stream $\mathcal{S} = (q_1, q_2, \dots, q_t, \dots)$, where $q_t$ is the query arriving at the LLM gateway at global time step $t$. Under the \textit{untraceability assumption}, the gateway observes only the raw semantic content of $q_t$, devoid of any reliable user metadata (e.g., session IDs or IP addresses).

The adversary covertly injects the fragment set $\{x_1, \dots, x_n\}$ into $\mathcal{S}$. This results in an unknown, strictly increasing timestamp sequence $t_1 < t_2 < \dots < t_n$ such that $q_{t_i} = x_i$. Crucially, between any two attack fragments $q_{t_i}$ and $q_{t_{i+1}}$, the stream is interleaved with an arbitrary number of background queries originating from other independent users.

\noindent\textbf{Defense Objective.}
Let $\mathcal{H}_{t-1} = \{q_1, \dots, q_{t-1}\}$ denote the global history of all queries processed prior to time $t$. In contrast to a stateless oracle, a \textit{stateful defense} conditions its adjudication on prior interactions. Formally, our objective is to learn a stateful decision function $\mathcal{F}_{def}(q_t, \mathcal{H}_{t-1}) \to \{0, 1\}$ that incorporates temporal memory to intercept distributed malicious intents.

\subsection{Dataset Construction}
\label{sec:dataset}

\noindent \textbf{Motivation and Overview.}
The development of effective stateful defenses against decompositional jailbreaks necessitates a large-scale, high-quality dataset that accurately reflects the complexity of real-world traffic. While pioneering efforts such as DecomposedHarm \cite{yuehhan2025monitoringdecompositionattacksllms} have laid the groundwork, they are limited in scale and diversity, which is insufficient for training data-intensive models. 

\begin{table}[t]
\centering
\caption{Comparison between our proposed dataset and DecomposedHarm~\cite{yuehhan2025monitoringdecompositionattacksllms}, an in-session jailbreak dataset.}
\label{tab:dataset_comparison}
\begin{tabular}{lcc}
\toprule
\textbf{Feature} & \textbf{Ours} & \textbf{DecomposedHarm \cite{yuehhan2025monitoringdecompositionattacksllms}} \\
\midrule
Total Intents & 250,625 & 4,641 \\
Background Noise & 603,842 & 0 (No Isolated Requests)\\
Max Splits per Sample & 3 & 1 \\
Source Datasets & 12 & 3 \\
\bottomrule
\end{tabular}
\end{table}

To address the limitations of existing benchmarks, we construct a large-scale, comprehensive dataset that serves as the empirical foundation. As summarized in Table~\ref{tab:dataset_comparison}, our dataset substantially expands the total number of intents, incorporates realistic background noise, increases fragmentation complexity (i.e., the number of splits per sample), and draws from a broader range of source corpora. Our dataset conforms to a stricter threat model in which attack instructions are entirely stateless, requiring neither shared conversational context nor strict sequential ordering, thereby posing a rigorous challenge for detection and defense systems.

The dataset comprises 603,842 independent benign queries, 250,625 benign intents with their corresponding decompositions, and 8,681 malicious intents with their associated fragments. Data are partitioned into training, validation, and test sets following an 8:1:1 ratio. To preclude data leakage, partitioning is performed strictly at the intent level, ensuring that all fragments derived from the same intent reside within the same split, and that the model is evaluated exclusively on intents unseen during training.

\noindent \textbf{Unfragmented Benign Requests.}
To model the broad distribution of safe user interactions, we curate benign queries sourced from five high-quality instruction datasets: CodeAlpaca~\cite{codealpaca}, Dolly~\cite{DatabricksBlog2023DollyV2}, LMSYS-Chat-1M~\cite{zheng2024lmsyschat1mlargescalerealworldllm}, UltraChat~\cite{ding2023enhancingchatlanguagemodels}, and WizardLM~\cite{xu2025wizardlmempoweringlargepretrained}.

Our curation pipeline prioritizes both safety and semantic distinctness. We first exclude all instructions flagged as potentially unsafe in their source datasets. The remaining candidates are then subjected to an additional safety audit via Llama-3-8B-Guard~\cite{inan2023llamaguardllmbasedinputoutput}, and any query failing the guardrail check is discarded. To reduce duplication within a dataset and across datasets, we apply a three-stage deduplication pipeline: (1) exact string matching, (2) MinHash-based similarity filtering, and (3) $n$-gram overlap analysis.

\noindent \textbf{Fragmented Benign Intents.}
To faithfully emulate legitimate users performing cross-session interactions, we extract benign intents from three complex task datasets: Stanford Alpaca~\cite{alpaca}, AlpacaEval~\cite{dubois2025lengthcontrolledalpacaevalsimpleway}, and Magpie~\cite{xu2024magpiealignmentdatasynthesis}. These intents are subject to the same safety filtration pipeline applied to unfragmented queries, combining metadata filtering with Llama-3-8B-Guard~\cite{inan2023llamaguardllmbasedinputoutput} verification, followed by the full three-stage deduplication procedure.

Verified benign intents are subsequently decomposed using three distinct models: \textit{Mistral-Small-24B-Instruct-2501-abliterated}~\cite{jiang2023mistral7b}, \textit{Qwen3-32B-abliterated}~\cite{yang2025qwen3technicalreport}, and \textit{Qwen3-30B-A3B-abliterated}, all of which are orthogonalized variants produced via refusal direction ablation~\cite{arditi2024refusallanguagemodelsmediated}. Crucially, these models are identical to those employed for malicious decomposition. Sharing decomposition models across both benign and malicious data is a deliberate design choice intended to avoid superficial stylistic artifacts introduced by different generation processes.

\noindent \textbf{Fragmented Malicious Intents.}\label{sec:dataset_malicious}
To construct a comprehensive set of adversarial attack scenarios, we collect initial malicious intents from four widely adopted safety evaluation benchmarks: HarmBench~\cite{mazeika2024harmbenchstandardizedevaluationframework}, AdvBench~\cite{zou2023universaltransferableadversarialattacks}, StrongReject~\cite{souly2024strongrejectjailbreaks}, and BeaverTails~\cite{ji2023beavertailsimprovedsafetyalignment}. These intents are processed through the same three-stage deduplication pipeline (exact match, MinHash, $n$-gram overlap) to ensure a semantically diverse collection of attack objectives.

Decomposition is carried out using the CKA-Agent framework~\cite{wei2025trojanknowledgebypassingcommercial} with the same three abliterated splitters as the benign pipeline, whose refusal direction has been ablated~\cite{arditi2024refusallanguagemodelsmediated} so that they do not reject adversarial decomposition prompts. We select CKA-Agent primarily because it integrates an automated verification mechanism capable of evaluating both the overall success of the reconstructed attack and whether each isolated sub-query is deemed safe by the target model. To guarantee both the efficacy and stealthiness of the resulting attacks, we impose a strict quality criterion: an intent-decomposition pair is retained in the dataset only if the aggregated response is assigned the highest rating, indicating that the output is both harmful and actionable. This constraint ensures that the decomposed fragment sequences successfully circumvent single-query defenses while collectively fulfilling the adversary's malicious objective. All generated sub-queries are finally subjected to the three-stage global deduplication pipeline to maintain semantic uniqueness across the full dataset. 

\subsection{Overview of TwinGate}
\label{sec:twingate_overview}

\begin{figure*}[t]
  \centering
  \includegraphics[width=1.0\linewidth]{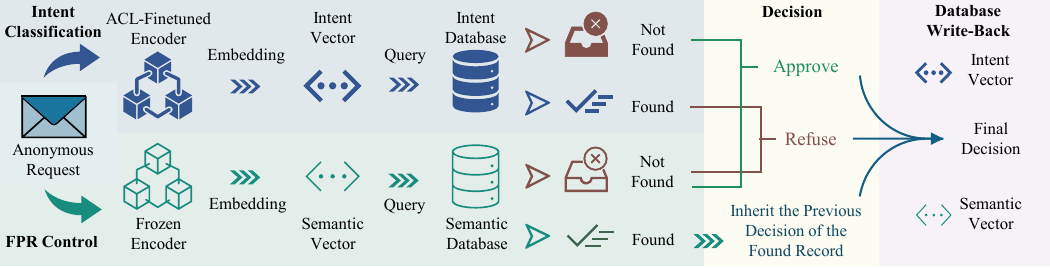}
  \caption{The end-to-end workflow of TwinGate. For each incoming request, the system performs dual encoding, stateful vector querying, and decision making (including decision inheritance and ACL intent clustering), followed by an asynchronous write-back to the vector database.}
  \Description{}
  \label{fig:twingate_workflow}
\end{figure*}

Figure~\ref{fig:twingate_workflow} illustrates the end-to-end TwinGate workflow. The design factors the defense into an offline training phase (Section~\ref{sec:acl}), which shapes an intent-clustering latent space, and an online deployment phase (Section~\ref{sec:decision_logic}), which adjudicates a continuous request stream against a globally maintained vector repository.

\noindent \textbf{System Architecture.}
TwinGate comprises four logical components that together realize the stateful decision function $\mathcal{F}_{def}(q_t, \mathcal{H}_{t-1})$. (i) A \textit{frozen encoder} produces a semantic vector capturing the literal surface meaning of $q_t$. (ii) An \textit{ACL-finetuned encoder} produces an intent vector that maps $q_t$ into a latent space organized by malicious objective rather than topical similarity. (iii) A \textit{global vector repository} stores both representations of every previously processed query in $\mathcal{H}_{t-1}$, and is queried in parallel by both encoders. (iv) A \textit{two-stage decision module} first inherits a prior verdict via a fast-path bypass when the semantic vector closely matches a historical entry; otherwise it adjudicates $q_t$ against the intent-vector database to detect membership in an established malicious cluster. After the verdict is returned to the gateway, both vectors and the decision are asynchronously written back to the repository, sustaining the temporal memory required by $\mathcal{F}_{def}$ without blocking inference.

\noindent \textbf{Decoupled Stateful Retrieval.}
Concatenating historical requests inside an LLM context window is untenable under high-throughput, untraceable traffic: interleaved queries rapidly exhaust the window, and Transformer~\cite{10.5555/3295222.3295349} attention scales quadratically. Offloading $\mathcal{H}_{t-1}$ to an external vector repository replaces in-context reasoning with retrieval, reducing the per-request cost to two lightweight encoder passes plus a top-1 similarity query and decoupling memory growth from inference latency.

\noindent \textbf{Intent Clustering via Asymmetric Contrastive Learning.}
Standard pre-trained encoders organize representations by surface similarity, which is insufficient when a unified malicious intent is deliberately fragmented into semantically disparate queries. ACL reorients the latent space around intent: fragments derived from a common malicious objective are pulled into compact clusters, while benign queries act exclusively as repulsive negatives so the broad benign manifold remains structurally intact. The training objective is formalized in Section~\ref{sec:acl}.

\noindent \textbf{Decision Inheritance for FPR Control.}
Aggressive intent clustering inevitably elevates the false-positive risk for repeated or near-duplicate benign requests, whose embeddings may drift toward established malicious clusters by virtue of topical proximity alone. The frozen encoder addresses this via a high-precision repetition check: when an incoming query closely matches a previously adjudicated request in semantic space, the system inherits the prior verdict on a fast path that bypasses ACL evaluation entirely, anchoring the false positive rate (FPR) to the small set of admitted historical decisions. The complete two-stage routing is formalized in Section~\ref{sec:decision_logic}.

\subsection{Asymmetric Contrastive Learning}
\label{sec:acl}

Section~\ref{sec:twingate_overview} positioned the ACL-finetuned encoder as TwinGate's intent-clustering component; this subsection formalizes its offline training over a DeBERTaV3~\cite{he2023debertav3improvingdebertausing} backbone. Diverse decomposition strategies fragment a single malicious objective into queries appearing semantically unrelated, so contrastive objectives optimized for surface similarity cannot recover the shared intent. ACL instead attracts fragments of a common malicious objective while using benign traffic as a repulsive boundary, yielding a geometry discriminative against attacks yet preserving legitimate query structure.

\noindent \textbf{Architecture and Representations.}
We employ an encoder-only architecture comprising a pre-trained Transformer backbone and a 2-layer non-linear MLP projection head (Linear $\to$ GELU~\cite{hendrycks2023gaussianerrorlinearunits} $\to$ Linear). For a given input query $x$, the encoder extracts the contextualized `[CLS]` token embedding, which is then projected into a lower-dimensional latent space and $L_2$-normalized. We denote this normalized latent representation as $z = f_{\theta}(x) \in \mathbb{R}^d$, where $\theta$ represents the trainable parameters.

\noindent \textbf{Intent-Centric Labeling.}
Crucially, our labeling strategy completely discards the source of the decomposition. For any malicious intent $I_{mal} \in \mathcal{I}_{mal}$, let $\mathcal{X}(I_{mal})$ denote the set containing the original intent itself along with all its decomposed fragments generated across multiple splitters. We assign a uniform contrastive label $y_{mal}$ to all $x \in \mathcal{X}(I_{mal})$. This aggressive grouping forces the model to ignore the superficial generation artifacts of different splitters and focus entirely on the core malicious objective.

\noindent \textbf{Asymmetric Loss Formulation.}
We adapt the Supervised Contrastive Learning (SupCon) framework. A symmetric SupCon would tightly cluster benign queries as well, collapsing their natural manifold (e.g., coding, writing, casual chat) and pulling them dangerously close to malicious ones. ACL therefore introduces an asymmetry: only \textit{malicious} queries serve as anchors and exert positive attraction; benign queries act exclusively as \textit{negatives}, pushing malicious clusters away from the benign distribution.

Formally, consider a training batch $\mathcal{B}$ consisting of a set of malicious queries $\mathcal{B}_{mal}$ and benign queries $\mathcal{B}_{ben}$. For a malicious anchor $x_i \in \mathcal{B}_{mal}$, let $P(i) = \{p \in \mathcal{B} \setminus \{i\} \mid y_p = y_i\}$ be the set of positive indices (other fragments sharing the same malicious intent $I_{mal}$). The ACL loss for the anchor $x_i$ is defined as:
\begin{equation}
    \mathcal{L}_{i} = \frac{-1}{|P(i)|} \sum_{p \in P(i)} \log \frac{\exp(z_i \cdot z_p / \tau)}{\sum_{j \in \mathcal{B} \setminus \{i\}} \exp(z_i \cdot z_j / \tau)},
\end{equation}
where $\tau$ is a scalar temperature parameter regulating the concentration of the distribution.

The total batch loss is then computed exclusively over the malicious anchors:
\begin{equation}
    \mathcal{L}_{ACL} = \frac{1}{|\mathcal{B}_{mal}|} \sum_{i \in \mathcal{B}_{mal}} \mathcal{L}_{i}.
\end{equation}

In this formulation, if $x_k \in \mathcal{B}_{ben}$ is a benign query, it never serves as an anchor $i$, nor does it ever belong to a positive set $P(i)$. However, it actively participates in the denominator $\sum_{j \in \mathcal{B} \setminus \{i\}}$, acting as a universal negative. This asymmetric dynamic guarantees that fragments of the same malicious intent converge into dense, isolated clusters in the latent space, while the broad representation of benign traffic remains structurally intact and safely distanced from attack vectors.

\subsection{Dual-Encoder Decision Logic}
\label{sec:decision_logic}

\noindent \textbf{Online Deployment Phase.} Pairing the ACL-finetuned encoder of Section~\ref{sec:acl} with the frozen semantic encoder introduced in Section~\ref{sec:twingate_overview}, TwinGate's runtime, depicted in Figure~\ref{fig:twingate_workflow}, processes the continuous, untraceable request stream $\mathcal{S}$ via a two-stage retrieval and decision mechanism. For each incoming query $q_t$ at time $t$, the system extracts two distinct $L_2$-normalized representations:
\begin{enumerate}
    \item \textbf{Semantic Vector} $s_t$: Extracted via the frozen encoder. This vector accurately captures the literal and surface-level semantic meaning of the text.
    \item \textbf{Intent Vector} $z_t$: Extracted via the ACL-finetuned encoder ($z_t = f_{\theta}(q_t)$). This vector maps the query into the optimized intent-clustering latent space.
\end{enumerate}

Given the global history of previously processed queries $\mathcal{H}_{t-1}$, and two predefined similarity thresholds, the semantic threshold $\tau_{sem}$ and the intent threshold $\tau_{int}$, the stateful decision function $\mathcal{F}_{def}(q_t, \mathcal{H}_{t-1})$ operates through the following sequence:

\noindent \textbf{Stage 1: Semantic Equivalence Inheritance. }
The aggressive clustering nature of the ACL objective inherently elevates the risk of False Positives. If a user repeatedly submits variations of the same safe query, the ACL encoder might cluster them so densely that they inadvertently trigger a malicious threshold. To counteract this, TwinGate first evaluates historical semantic repetition. We compute the maximum semantic similarity between the current query and the history:
\begin{equation}
    m_{sem} = \max_{q_j \in \mathcal{H}_{t-1}} (s_t \cdot s_j).
\end{equation}
If $m_{sem} > \tau_{sem}$, the system identifies a historical query $q_{t^*}$ that is semantically equivalent to $q_t$. In this scenario, the system triggers a short-circuit logic: it bypasses the ACL evaluation entirely and directly inherits the final safety decision previously assigned to $q_{t^*}$. Formally, $\mathcal{F}_{def}(q_t, \mathcal{H}_{t-1}) = \mathcal{F}_{def}(q_{t^*}, \mathcal{H}_{t^*-1})$. This inheritance mechanism stabilizes the False Positive Rate by ensuring that repetitive benign traffic is handled consistently based on its initial, safe evaluation.

\noindent \textbf{Stage 2: Intent-based Attack Detection. }
If $m_{sem} \le \tau_{sem}$, $q_t$ is semantically novel and TwinGate queries the database with $z_t$ for the maximum historical intent similarity:
\begin{equation}
    m_{int} = \max_{q_j \in \mathcal{H}_{t-1}} (z_t \cdot z_j).
\end{equation}
If $m_{int} > \tau_{int}$, it indicates that $q_t$ belongs to an established cluster of malicious fragments traversing the system. Consequently, the defense mechanism intercepts the request, yielding $\mathcal{F}_{def}(q_t, \mathcal{H}_{t-1}) = 0$. Conversely, if the query remains isolated in the latent space ($m_{int} \le \tau_{int}$), it is deemed safe and permitted to reach the LLM, yielding $\mathcal{F}_{def}(q_t, \mathcal{H}_{t-1}) = 1$.

By cascading these two encoders, TwinGate effectively shields the aggressive malicious intent clustering (Stage 2) with a conservative semantic equivalence filter (Stage 1), achieving a robust balance between attack recall and benign traffic availability.

\subsection{Theoretical Bounds on Adaptive Attacks}
\label{sec:theoretical_bounds}

Sections~\ref{sec:twingate_overview}--\ref{sec:decision_logic} described how TwinGate operates; we now bound its robustness against a white-box adaptive adversary via a \textit{geometric} ceiling on fragments that fit beneath the detection threshold (Theorem~1) and a \textit{computational} ceiling on the discrete-search effort required to find them (Theorem~2).

As defined in Section~\ref{sec:threat_model}, the first fragment $x_1$ evades detection due to lack of prior context $\mathcal{H}_{t_1-1}$. For $i \ge 2$, evasion requires satisfying three constraints in the latent manifold $\mathcal{V}(I_{mal})$: (1) \textit{semantic novelty} $\max_{j < i} (s_{t_i} \cdot s_{t_j}) \le \tau_{sem}$, blocking fast-path inheritance; (2) \textit{intent evasion} $\max_{j < i} (z_{t_i} \cdot z_{t_j}) \le \tau_{int}$, blocking intent-cluster recall; and (3) \textit{effectiveness within $\mathcal{V}(I_{mal})$}, since random text satisfies (1)--(2) but cannot advance $I_{mal}$. Let $r = \arccos(\tau_{int})$ be the minimum angular distance between any two evasive intent embeddings imposed by constraint~(2), and model the local intent space as a $d_{int}$-dimensional sub-manifold ($d_{int} \le d-1$ from $L_2$ normalization) of maximum angular radius $R_{mal}$ centered at $I_{mal}$.

\noindent \textbf{Theorem 1 (Maximum Decomposition Limit).}
\textit{In the small-angle regime $r \ll 1$, the number of evasive fragments $n_{max}$ that an adversary can inject within $\mathcal{V}(I_{mal})$ is bounded by the spherical packing limit:}
\begin{equation}
    n_{max} \le \left( \frac{2 R_{mal}}{\arccos(\tau_{int})} \right)^{d_{int}}.
\end{equation}
\textit{Proof sketch:} Constraint~(2) forces any two evasive intent embeddings to lie at angular distance $\ge r$, so each occupies a non-overlapping spherical cap of radius $r/2$ centered on $z_{t_i}$. Packing such caps into $\mathcal{V}(I_{mal})$ of radius $R_{mal}$ yields the bound via the angular-volume ratio $\mathrm{vol}(R_{mal})/\mathrm{vol}(r/2) \approx (2R_{mal}/r)^{d_{int}}$. ACL training monotonically tightens the bound by compressing $R_{mal}$ via the same-intent attraction term, but the slack against constraint~(2) alone permits $n_{max} > 1$ whenever $R_{mal} > r/2$, consistent with the residual evasion observed in our adaptive-attack experiments (Section~\ref{sec:adaptive_attacks}).

\noindent \textbf{Theorem 2 (Computational Intractability of Discrete Evasion).}
The adversary operates in a discrete token space $\mathcal{T}^L$ where the encoder $f_\theta: \mathcal{T}^L \to \mathbb{S}^{d-1}$ is non-convex, so token-level perturbations induce discrete, lower-bounded jumps in latent space rather than infinitesimal shifts. Let $\mathcal{X}_{val} \subset \mathcal{T}^L$ be the finite set of token sequences preserving semantic validity and evasiveness for $I_{mal}$. To inject the $i$-th fragment, the adversary must satisfy $i-1$ historical exclusion constraints:
\begin{equation}
    \forall j \in \{1, \dots, i-1\}, \quad f_\theta(x_i) \cdot z_{t_j} \le \tau_{int}.
\end{equation}

Let $\gamma \in (0, 1)$ denote the fraction of $\mathcal{X}_{val}$ excluded by a single such constraint, and assume valid token sequences map approximately uniformly over $\mathcal{V}(I_{mal})$ so that successive constraints are quasi-independent. The feasible set then shrinks multiplicatively:
\begin{equation}
    |\mathcal{X}_{feas}^{(i)}| \le |\mathcal{X}_{val}| (1 - \gamma)^{i-1}.
\end{equation}

Consequently, the probability that a randomly drawn candidate satisfies all $i-1$ constraints scales as $(1-\gamma)^{i-1}$, and any local-search algorithm (e.g., CKA-Agent) requires:
\begin{equation}
    \mathbb{E}[Q_i] = \Omega\left( \frac{1}{(1 - \gamma)^{i-1}} \right).
\end{equation}

As $i$ grows, $\mathbb{E}[Q_i]$ scales exponentially, rendering fine-grained decompositional jailbreaks computationally intractable long before reaching the geometric packing limit of Theorem~1, a prediction validated empirically against white-box adaptive attackers in Section~\ref{sec:adaptive_attacks}.
\section{Hardware-Aware System Implementation}
\label{sec:implementation}

The practical deployment of TwinGate necessitates addressing both the semantic complexities of intent clustering and the rigorous latency requirements of production-grade LLM infrastructures. This section details the training dynamics, the theoretical and empirical justification for our stateful retrieval strategy, and our hardware-accelerated pipeline.

\subsection{Training Dynamics}
\label{sec:training_dynamics}

\noindent \textbf{The Semantic Gap and Triangle Inequality.} 
Clustering decomposed attacks is fundamentally challenged by the extreme semantic divergence between fragments derived from the same malicious intent. While a model may associate a specific fragment (e.g., \textit{``write a script to traverse folders''}) with its latent malicious objective (\textit{ransomware development}), establishing a direct link between two such disparate fragments is non-trivial. To bridge this gap, we incorporate the original, complete malicious intent into the training set as a \textit{dense latent anchor}. By assigning identical contrastive labels to the original intent and its constituent fragments, the Asymmetric Contrastive Learning (ACL) objective forces fragments to converge toward this anchor. Consequently, disparate fragments are brought into proximity within the manifold, constrained by the vector space triangle inequality.

\begin{figure}[t]
  \centering
  \includegraphics[width=1.0\linewidth]{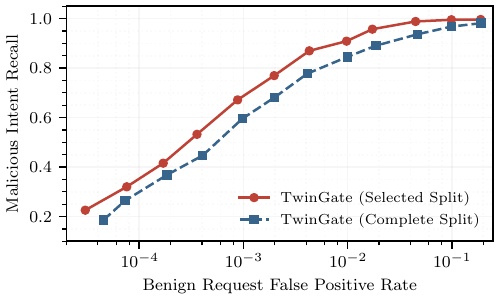}
  \caption{Impact of semantic pruning on ACL performance: The pruned dataset yields a superior Recall-FPR curve by eliminating latent space collisions.}
  \Description{Comparison of Recall FPR trade off curves.}
  \label{fig:train_consistency}
\end{figure}

\noindent \textbf{Semantic Pruning for Training Consistency.} 
To maintain alignment between training and inference distributions, we employ semantic pruning. Standard character-level deduplication fails to identify sophisticated semantic equivalents. In our deployment architecture (Section~\ref{sec:decision_logic}), the frozen encoder's short-circuit logic intercepts these duplicates before they reach the ACL evaluation. Retaining such samples during ACL training introduces severe gradient conflicts: the ACL objective would attempt to forcefully separate semantically proximate samples that belong to different intent clusters, destabilizing the manifold. By proactively pruning samples with high cosine similarity in the frozen encoder's latent space, we eliminate these collisions. As illustrated in Figure~\ref{fig:train_consistency}, this pruning strategy significantly optimizes the Recall-FPR trade-off by mitigating destructive gradient updates.

\noindent \textbf{Contrastive Batch Construction.} 
The efficacy of the Supervised Contrastive (SupCon) loss relies heavily on batch composition. Each training batch is constructed to contain fragments from at least two distinct malicious intents alongside benign queries. This diversity prevents the model from trivially collapsing all malicious inputs into a single undifferentiated cluster, instead compelling it to differentiate between specific malicious objectives. We define an epoch as a complete pass over the malicious samples and train TwinGate for three epochs on a cluster of $8 \times$ NVIDIA A100 (40GB) GPUs.

\subsection{Threshold Calibration and the \texorpdfstring{$k=1$}{k=1} Strategy}
\label{sec:thresholds}

\noindent \textbf{Threshold Selection via 2D Grid Search.} 
To ensure generalizability, threshold calibration is strictly decoupled from the final evaluation. We perform a two-dimensional grid search over the semantic threshold $\tau_{sem}$ and intent threshold $\tau_{int}$ using an isolated validation set. The optimal pair, determined by maximizing recall under a strict FPR constraint, is subsequently frozen. All metrics reported in Section~\ref{sec:evaluation} are derived from the unseen test set using these frozen parameters to guarantee a zero-shot evaluation protocol with no data leakage.

\noindent \textbf{The Rationale for $k=1$.} 
The parameter $k$ represents the minimum number of historical requests in $\mathcal{H}_{t-1}$ required to exceed $\tau_{int}$ to execute an interception. While a $k=0$ policy represents a stateless filter incapable of detecting decomposition, $k=1$ represents the most responsive stateful policy. We argue that $k=1$ is optimal for two primary reasons:
\begin{enumerate}
    \item \textbf{Security Constraints:} Increasing $k$ inherently allows an adversary to successfully extract multiple harmful responses before the system intervenes, exacerbating the risk of malicious information leakage.
    \item \textbf{Scalability Constraints:} A $k=1$ threshold reduces the defense mechanism to a highly optimized top-1 proximity query. Conversely, requiring $k > 1$ necessitates maintaining complex, stateful session counters across millions of anonymous users, imposing prohibitive computational and memory overhead.
\end{enumerate}

\noindent \textbf{Empirical Validation.} 
To validate this architectural decision, we evaluated performance on a stringent test subset containing only one trajectory per intent. As shown in Table~\ref{tab:k_ablation}, increasing $k$ from 1 to 2 reduces recall from 0.77 to 0.73. More critically, $k=1$ minimizes the \textit{First Intercept Position} to 2.65, meaning the attacker extracts only $\approx 1.65$ benign-looking responses before the entire attack vector is blocked. At $k=3$, this value rises to 3.11, significantly widening the system's window of vulnerability.

\begin{table}[t]
  \centering
  \caption{Impact of parameter $k$ on detection performance (FPR $< 2 \times 10^{-3}$).}
  \label{tab:k_ablation}
  \small
  \begin{tabular*}{\linewidth}{@{\extracolsep{\fill}}lccc}
    \toprule
    \textbf{Required Neighbors ($k$)} & \textbf{1} & \textbf{2} & \textbf{3} \\
    \midrule
    \textbf{Malicious Intent Recall} & 0.77 & 0.73 & 0.51 \\
    \textbf{Avg. First Intercept Position} & 2.65 & 3.40 & 4.11 \\
    \bottomrule
  \end{tabular*}
\end{table}

\subsection{Hardware-Aware Pipeline}
\label{sec:hardware_pipeline}

To ensure seamless integration into production environments, TwinGate utilizes a hardware-accelerated, four-GPU pipelined topology designed to sustain high throughput and ultra-low latency. This pipeline is deployed on a single server with $4 \times$ NVIDIA A100 (40GB) GPUs interconnected via NVLink.

\noindent \textbf{Asynchronous Dispatch and Parallel Extraction.} 
The host CPU serves as a lightweight gateway limited to request arrival handling, tokenization, and dynamic batching. To prevent execution bottlenecks, feature extraction is strictly parallelized: GPU 0 hosts the frozen encoder for semantic vectors ($s_t$), while GPU 1 runs the ACL-finetuned encoder for intent vectors ($z_t$). This asynchronous dispatch decouples I/O operations from heavy neural computation.

\noindent \textbf{Zero-Copy Routing and In-Memory Database.} 
Routing intermediate tensors through the host CPU creates PCIe bottlenecks. TwinGate circumvents this using NVLink for zero-copy, point-to-point transfer of $L_2$-normalized vectors directly to the database GPUs (GPU 2 and GPU 3). The global historical repository $\mathcal{H}_{t-1}$ is maintained entirely in FP16 High Bandwidth Memory (HBM). To prevent memory exhaustion when computing dense cosine similarities over millions of vectors, the database dynamically profiles VRAM and adaptively chunks the historical matrix to maximize Tensor Core utilization.

\noindent \textbf{On-GPU Decision Logic.} 
The two-stage decision logic executes entirely within VRAM. GPU 2 computes $m_{sem}$ and transmits a fast-path flag via NVLink to GPU 3. GPU 3 then aggregates this signal with its own $m_{int}$ computation to yield the final binary safety decision $\mathcal{F}_{def}$ before returning it to the CPU. Concurrently, new request vectors and decisions are inserted into the HBM databases via a background CUDA stream, ensuring that these $O(1)$ updates remain completely hidden from the critical path of incoming requests.
\section{Experimental Evaluation}
\label{sec:evaluation}

In this section, we present a comprehensive empirical evaluation of TwinGate, systematically assessing its defensive efficacy against decompositional jailbreaks and its system-level performance. Our evaluation is designed to answer whether TwinGate can provide robust, stateful security guarantees without imposing prohibitive latency or throughput overheads.

\subsection{Main Security Effectiveness}
\label{subsec:main_results}

Defending against decompositional jailbreaks fundamentally requires stateful intent identification, distinguishing it from traditional static filtering mechanisms. We begin by analyzing the fundamental trade-off between security stringency and usability.

\noindent \textbf{Key Evaluation Metrics.} To rigorously quantify the effectiveness of TwinGate, we define two principal metrics based on the notation established in Section~\ref{sec:threat_model}:

\begin{itemize}
    \item \textbf{Malicious Intent Recall (Recall):} This metric measures the proportion of malicious intents where at least one constituent request (slice) is correctly intercepted. Formally, given a set of malicious decompositional jailbreaks $\mathcal{A} = \{I_1, I_2, \dots, I_n\}$, where each intent $I_i$ is decomposed into a sequence of slices $\{s_{i,1}, \dots, s_{i,m_i}\}$, the recall is defined as:
    \[ \text{Recall} = \frac{|\{I_i \in \mathcal{A} \mid \exists s_{i,j} : \mathcal{D}(s_{i,j}) = \text{Malicious}\}|}{n} \]
    Recall captures the system's overarching capability to disrupt the execution chain of a decompositional jailbreak before the complete malicious payload is reconstructed by the backend model.

    \item \textbf{Benign Request False Positive Rate (FPR):} This metric quantifies the proportion of independent benign requests, or slices belonging to legitimate user intents, that are erroneously flagged as malicious. A high FPR severely degrades the user experience by interrupting valid conversational workflows. Formally:
    \[ \text{FPR} = \frac{|\{s \in \mathcal{S}_{benign} \mid \mathcal{D}(s) = \text{Malicious}\}|}{|\mathcal{S}_{benign}|} \]
\end{itemize}

It is imperative to evaluate the \textit{trade-off} between Recall and FPR, rather than observing them in isolation. A practical security system must sustain a high Recall while strictly bounding the FPR to a negligible threshold (e.g., $<0.2\%$) to preserve the system's utility for normal interactions. Before detailing these results, we emphasize a crucial detail of our experimental setup: while the original monolithic malicious intent is utilized as an anchor during the ACL training phase, it is strictly excluded from the test request streams. This faithfully replicates the real-world scenario where the target LLM never observes the complete malicious intent.

\noindent \textbf{Comparison with Baselines.} We benchmark TwinGate against a representative set of state-of-the-art defense mechanisms:
\begin{itemize}
    \item \textbf{Llama-Guard-3-8B}: A widely adopted, safety-aligned guardrail model \cite{grattafiori2024llama3herdmodels} designed to filter single instructions in isolation without conversational memory.
    \item \textbf{Intent-FT}: A baseline model fine-tuned explicitly for intent recognition \cite{zhang2024intentionanalysismakesllms}, functioning as an enhanced single-turn semantic filter.
    \item \textbf{Window Monitor}: A sequential monitoring approach that maintains a fixed history window to detect temporally distributed suspicious patterns \cite{yuehhan2025monitoringdecompositionattacksllms}.
\end{itemize}

\begin{figure}[t]
  \centering
  \includegraphics[width=1.0\linewidth]{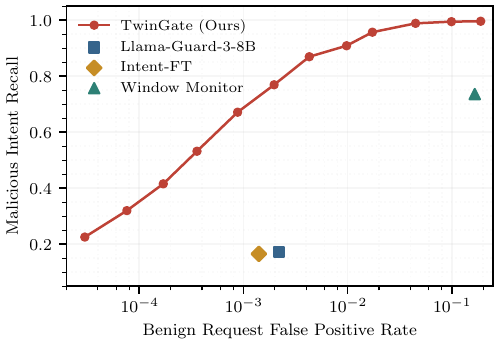}
  \caption{Recall-FPR trade-off curve of TwinGate compared with three baseline methods (Llama-Guard-3-8B, Intent-FT, and Window Monitor) represented as scatter points.}
  \Description{}
  \label{fig:recall_fpr}
\end{figure}

Owing to the threshold calibration described in Section~\ref{sec:thresholds}, TwinGate produces a continuous Recall-FPR curve, whereas static baselines report only individual scatter points. As illustrated in Figure~\ref{fig:recall_fpr}, TwinGate comprehensively outperforms all three baseline methods across the entire operational spectrum. Specifically, Llama-Guard-3-8B and Intent-FT exhibit similar performance, but their recall is significantly lower than that of TwinGate at an equivalent FPR. Meanwhile, Window Monitor achieves a relatively competitive recall, but suffers from an unacceptably high FPR, which is nearly 100 times that of TwinGate at the same recall level. In absolute terms, the FPR of Window Monitor exceeds 10\%, rendering it practically unusable in real-world deployment scenarios where benign traffic must not be excessively disrupted.

\subsection{Throughput and Latency Profiling}
\label{subsec:perf_results}

To validate the deployment feasibility of TwinGate, we conduct a rigorous system-level performance evaluation against the aforementioned baselines, focusing on the overhead introduced by the defense layer.

\noindent \textbf{Experimental Setup.} To ensure a strictly fair comparison, TwinGate runs in the four-GPU topology described in Section~\ref{sec:hardware_pipeline}, while baseline methods are deployed on the same server using the highly optimized vLLM~\cite{10.1145/3600006.3613165} serving framework with full access to all four GPUs for inference to maximize their potential throughput.

We model user request arrivals as a Poisson process with an arrival rate parameter $\lambda$. By progressively increasing $\lambda$ until the system reaches saturation, we identify the maximum sustainable throughput, where the stabilized queries per second (QPS) equals the arrival rate $\lambda$. For the main throughput and latency evaluation in this subsection, the vector database is pre-loaded with a history of 1 million requests to simulate a realistic operational state.

\noindent \textbf{Results and Analysis.} We focus on two critical performance metrics: \textbf{$P_{99}$ Latency} and \textbf{Throughput (QPS)}. $P_{99}$ latency is paramount for user experience: in a typical Prefill-Decoding (P-D) disaggregated LLM serving architecture, hiding the defense's latency within the backbone's prefill phase makes the security overhead transparent to end-users.

\begin{figure}[t]
  \centering
  \includegraphics[width=1.0\linewidth]{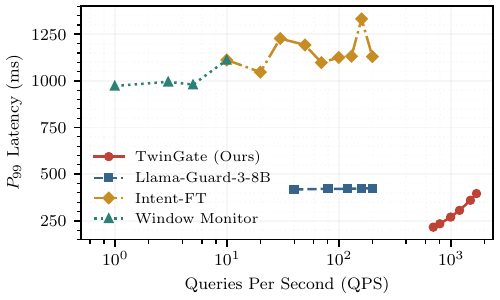}
  \caption{Comparison of $P_{99}$ latency versus throughput (QPS) for TwinGate and three baseline methods (Llama-Guard-3-8B, Intent-FT, and Window Monitor).}
  \Description{}
  \label{fig:p99_qps}
\end{figure}

As illustrated in Figure~\ref{fig:p99_qps}, our profiling results reveal that TwinGate comprehensively outperforms all three baseline methods. Specifically, TwinGate maintains a consistently low $P_{99}$ latency of $<300$ms even under severe load conditions. In contrast, Intent-FT and Window Monitor not only achieve a maximum QPS that is more than $10\times$ lower than TwinGate, but they also suffer from substantially higher latencies; in absolute terms, their $P_{99}$ latencies quickly approach $1000$ms, a delay that would significantly degrade the user experience in real-time interactions. Meanwhile, Llama-Guard-3-8B maintains a somewhat acceptable latency profile prior to saturation, yet it remains higher than that of TwinGate, and its maximum throughput is still an order of magnitude lower. The fundamental performance advantage of TwinGate derives from its drastically reduced computational footprint. While the baselines rely on heavy 7B-parameter class models, the dual-encoder components in TwinGate are lightweight (approximately 400M parameters) and require only two forward passes per request in the worst-case scenario. Consequently, despite dedicating only half of the available GPUs to active neural inference, the computational efficiency of the TwinGate architecture allows it to sustain a throughput exceeding $1700$ QPS. This represents nearly an order of magnitude improvement over the 200 QPS threshold characteristic of the 7B-class baseline filters, confirming that TwinGate provides robust stateful security without sacrificing system-level efficiency.

\subsection{Database Performance and Scalability}
\label{subsec:db_scalability}

Given that TwinGate relies on a stateful vector database to track cross-session interactions, evaluating the scalability of this component is critical. We stress-test the database retrieval subsystem using a fixed arrival rate of 1000 QPS, progressively increasing the database scale from 1 million to 7 million stored request vectors.

\begin{figure}[t]
  \centering
  \includegraphics[width=1.0\linewidth]{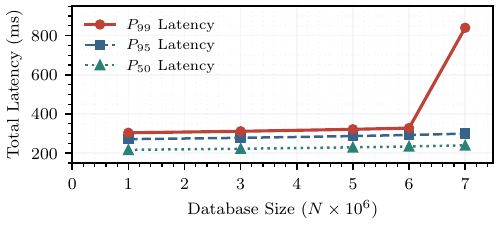}
  \caption{Latency ($P_{50}$, $P_{95}$, $P_{99}$) scaling with database size. The system maintains stable, sub-linear latency growth up to 6 million vectors, followed by a sharp performance degradation at 7 million due to VRAM exhaustion.}
  \Description{}
  \label{fig:db_scalability}
\end{figure}

\noindent \textbf{Latency Distribution \& Hardware Limits.} As illustrated in Figure~\ref{fig:db_scalability}, we plot the $P_{50}$, $P_{95}$, and $P_{99}$ latency curves as the database scales from 1 million to 7 million request vectors. Our measurements provide concrete evidence for three key operational characteristics of the TwinGate architecture. First, the system possesses a massive functional capacity: a single 40GB VRAM GPU can robustly accommodate up to 6 million historical requests for highly efficient, low-latency retrieval. Second, the end-to-end system latency is heavily dominated by the forward propagation of the two neural encoders, while the overhead introduced by the vector database itself is minimal. Third, within this 6-million-vector operational envelope, the database performance exhibits exceptional stability. The three latency curves remain remarkably flat, showing almost no noticeable degradation as the data scale expands from 1 million to 6 million. However, the stability holds only until the system hits a hard physical ceiling. At 7 million requests, the $P_{99}$ curve spikes dramatically to over 840~ms. This severe performance degradation signifies the exhaustion of the 40GB HBM capacity, leading to memory thrashing and indicating the absolute capacity limit of a single GPU.

\noindent \textbf{Bounded Memory and Slow-Loris Resilience.} Although our profiling validates that a single 40GB GPU can robustly cache 6 million historical requests, and industrialized vector databases (e.g., Milvus~\cite{10.1145/3448016.3457550} or Faiss~\cite{douze2025faisslibrary}) can seamlessly scale this capacity, accumulating vectors indefinitely is fundamentally impractical. It inevitably exhausts finite storage resources and violates privacy principles regarding the long-term retention of user data, necessitating a cache eviction policy. However, an adaptive adversary aware of such memory bounds will execute \textit{slow-loris} decomposition attacks, deliberately maximizing the temporal intervals between malicious fragments to outlive the system's eviction cycle. To evaluate TwinGate under this worst-case scenario, we construct a specialized adversarial request stream where fragments of the same malicious intent are spaced as far apart as possible by massive benign background traffic.

To counter this threat without inflating storage, we introduce a decoupled, clustering-based Least Recently Used (LRU) strategy. Instead of naive eviction, when a database reaches its capacity limit, historical requests with cosine distances exceeding a similarity threshold are merged into a single representative vector. The merged vector's timestamp is set to its constituents' latest match time and refreshed by every subsequent match, with the Semantic DB and Intent DB running this clustering-LRU independently.

\begin{figure}[t]
  \centering
  \includegraphics[width=1.0\linewidth]{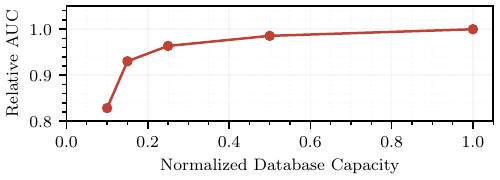}
  \caption{Relative AUC of the Recall-FPR curve across different normalized database capacities. The defense efficacy remains highly resilient down to a $25\%$ capacity ratio.}
  \Description{}
  \label{fig:lru_performance}
\end{figure}

We evaluate the defense efficacy by measuring the Area Under the Curve (AUC) of the Recall-FPR trade-off, normalized against an infinite-capacity baseline ($\text{AUC}_{rel} = 1.0$). As illustrated in Figure~\ref{fig:lru_performance}, by varying the capacity ratio $x$ (maximum database size / total requests), our results demonstrate that $\text{AUC}_{rel}$ remains highly resilient under aggressive compression. A noticeable performance decay only emerges when $x$ drops to $0.15$, followed by severe degradation at $0.10$. This confirms that TwinGate can sustain robust stateful defense against temporally-delayed attacks while retaining only $25\%$ of the traffic volume, thereby strictly bounding hardware costs and mitigating privacy risks.

\subsection{Robustness Against Adaptive Attacks}
\label{sec:adaptive_attacks}

While TwinGate demonstrates high efficacy against static datasets, its viability in deployment hinges on its resilience against an adaptive adversary actively attempting to circumvent the vector retrieval and dual-encoder logic. We adopt a strict white-box threat model, assuming the attacker has full knowledge of TwinGate's architecture, the ACL objective, the dual-encoder logic, and the neighbor count \texorpdfstring{$k=1$}{k=1}. Under this pessimistic assumption, the adversary may pursue two distinct objectives: (1) \textit{Evasion}, manipulating the semantic representation of their decomposed prompts to bypass the vector retrieval; and (2) \textit{False Positive Rate (FPR) Inflation}, deliberately polluting the system's memory to cause benign user requests to be falsely flagged.

\noindent \textbf{Evasion via Semantic Manipulation.}
To achieve evasion, a white-box attacker theoretically has two avenues: decreasing the similarity of their current malicious fragment to previously blocked fragments in the ACL database, or increasing its similarity to previously allowed benign requests to exploit the Fast-path (decision inheritance) mechanism for a free pass. To evaluate these strategies, we employ the state-of-the-art decomposition framework, CKA-Agent, as our base attacker. We measure the Attack Success Rate (ASR) against the maximum allowed attack attempts, which serves as a proxy for attack cost (scaling almost linearly with attempts).

\begin{figure}[t]
  \centering
  \includegraphics[width=1.0\linewidth]{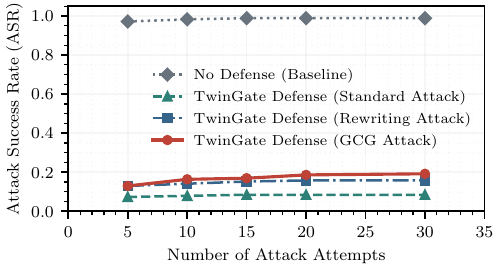}
  \caption{Attack Success Rate (ASR) versus the number of attack attempts under different threat configurations: No Defense, TwinGate Defense (Standard Attack), Rewriting Attack, and GCG Attack.}
  \Description{}
  \label{fig:asr_attempts}
\end{figure}

As illustrated in Figure~\ref{fig:asr_attempts}, we plot four distinct ASR curves against the number of attack attempts: an undefended system, a system defended by TwinGate (Standard Attack), a Rewriting Attack, and a powerful white-box GCG~\cite{zou2023universaltransferableadversarialattacks} Attack utilizing the first adaptive avenue (decreasing similarity). Without any defense, the ASR of the base CKA-Agent stabilizes at approximately 0.99. Upon deploying TwinGate, the ASR of the unmodified CKA-Agent plummets to roughly 0.09. Notably, this ASR is significantly lower than the theoretical upper bound implied by our main evaluation. This discrepancy arises from the interactive nature of CKA-Agent: in a live attack, the agent generates numerous intermediate failed attempts. These failures are also embedded and stored in the vector database, dynamically enriching the semantic cluster of the underlying malicious intent and making subsequent fragments even easier for TwinGate to intercept.

To evaluate the first adaptive avenue (decreasing similarity), we test two strategies. First, the attacker employs an auxiliary LLM to dynamically rewrite and paraphrase every fragment generated by CKA-Agent. As shown in the curves, this Rewriting strategy increases the ASR to around 0.16. Second, we consider a strict white-box threat model where the attacker possesses the exact weights of the ACL model. Under this setting, the attacker employs the Greedy Coordinate Gradient (GCG) algorithm to iteratively optimize adversarial suffixes for their subsequent requests, explicitly pushing their representation vectors as far away as possible from all previously blocked attack vectors in the database. Even under this extreme white-box optimization, the ASR peaks at only 0.18. Crucially, none of the four scenarios exhibit a meaningful ASR increase with more attempts: brute-force, paraphrasing, and gradient optimization all merely raise the attack cost (tokens and compute) without yielding evasion gains.

For the second adaptive avenue (increasing similarity to benign requests), we evaluate various benign-malicious blending strategies, where the attacker explicitly embeds the malicious intent within normal, safe contexts. As reported in Table~\ref{tab:obfuscation_asr}, these obfuscation techniques fail to bypass the defense, with the highest ASR reaching only $0.10$. Furthermore, because the inputted requests under these blending strategies include not only attacks but also benign text, we must ensure that such defensive isolation does not penalize normal usage. As shown in the newly appended FPR column of Table~\ref{tab:obfuscation_asr}, the FPR does not exhibit any significant increase, remaining stably low across all strategies. This demonstrates that TwinGate accurately identifies the underlying malicious intent without being misled by the injected benign context.

\begin{table}[t]
  \caption{Attack Success Rate (ASR) and False Positive Rate (FPR) under various benign-malicious blending strategies.}
  \label{tab:obfuscation_asr}
  \centering
  \small
  \begin{tabular}{llcc}
    \toprule
    \textbf{Blending Strategy} & \textbf{Position} & \textbf{ASR} & \textbf{FPR} \\
    \midrule
    1 Malicious + 1 Benign & First & 0.07 & $2.18 \times 10^{-3}$ \\
    1 Malicious + 1 Benign & Last & 0.04 & $2.16 \times 10^{-3}$ \\
    \midrule
    Fill Context (1 Malicious) & First & 0.10 & $1.90 \times 10^{-3}$ \\
    Fill Context (1 Malicious) & Random & 0.02 & $2.43 \times 10^{-3}$ \\
    Fill Context (1 Malicious) & Last & 0.08 & $1.86 \times 10^{-3}$ \\
    \bottomrule
  \end{tabular}
\end{table}

\noindent \textbf{FPR Inflation via Intent Pollution (Semantic DoS).}
Alternatively, an adversary may abandon evasion and instead aim to degrade system usability by intentionally inflating the False Positive Rate (FPR), effectively launching a semantic Denial-of-Service (DoS) attack. Theoretically, there are two primary strategies to pollute the vector database and cause benign requests to be rejected: (1) crafting a malicious request whose ACL representation is artificially pushed close to the cluster of future benign requests; or (2) crafting a request that closely resembles a benign query but contains adversarial triggers to force a false positive from the primary evaluation model, thereby poisoning the Fast-path cache for subsequent legitimate users.

To evaluate these threats under a unified and extremely pessimistic white-box setting, we design an experiment that maximizes the attacker's advantage. We assume the attacker not only possesses an exact copy of TwinGate's architecture, weights, and thresholds, but also has an oracle's knowledge of the exact benign instructions normal users will issue in the future. The attacker samples $n$ safe requests (comprising both independent queries and benign decomposition slices). Using the Greedy Coordinate Gradient (GCG) algorithm, the attacker appends adversarial suffixes to these requests, aiming to force the system to classify them as malicious. We further assume the attacker has full control over the request stream, ensuring these adversarial queries arrive and pollute the database just before the legitimate users send their clean versions. After this targeted poisoning phase, we re-execute the evaluation pipeline from Section~\ref{subsec:main_results}.

\begin{figure}[t]
  \centering
  \includegraphics[width=1.0\linewidth]{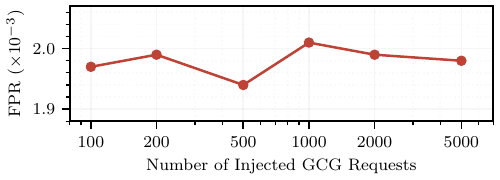}
  \caption{Impact of the number of GCG-poisoned injections on the False Positive Rate (FPR). The FPR remains stably around $2 \times 10^{-3}$ regardless of attack scale.}
  \Description{}
  \label{fig:fpr_pollution}
\end{figure}

As illustrated in Figure~\ref{fig:fpr_pollution}, the results reveal that the FPR does not exhibit any noticeable increasing trend as the number of injected GCG-poisoned samples $n$ grows. Instead, the FPR remains highly stable at approximately $2 \times 10^{-3}$, demonstrating that the system's usability is unaffected by the scale of the semantic DoS attack. The resilience against this pollution stems from the strict threshold calibration of the dual-encoder system. Even if the attacker successfully forces an adversarial sample to be misclassified and stored, the GCG-optimized suffix inherently alters the representation in the high-dimensional feature space. Because TwinGate operates with highly stringent similarity thresholds to prevent collateral damage, the subsequent clean, normal requests do not fall within the tight retrieval radius of the poisoned, adversarially-perturbed samples. Consequently, the Fast-path inheritance is not triggered, and the semantic DoS attack fails to materialize.

\subsection{Ablation Study}
\label{subsec:ablation}

\begin{figure}[t]
  \centering
  \includegraphics[width=1.0\linewidth]{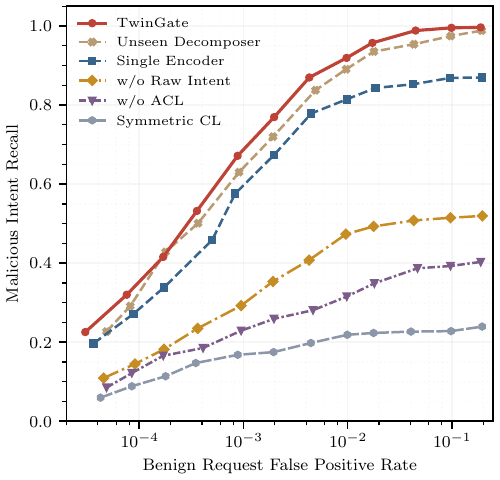}
  \caption{Recall-FPR curves for TwinGate against ablation variants (Single Encoder, w/o ACL, Symmetric CL, w/o Raw Intent) and an Unseen Decomposer.}
  \Description{}
  \label{fig:ablation}
\end{figure}

\noindent \textbf{The Necessity of the Dual-Encoder Architecture.}
To justify the retention of the frozen encoder, we investigate the system's performance when the Semantic Equivalence Inheritance mechanism (the frozen encoder) is removed, relying solely on the ACL encoder for all decisions. As shown in Figure~\ref{fig:ablation}, we evaluate this configuration across the Recall-FPR spectrum. While the full TwinGate system maintains robust performance, the \textit{Single Encoder} ablation exhibits a significant decline. It must be emphasized that this evaluation is performed on our standard test set, which was strictly deduplicated during dataset construction and therefore contains no repeated requests. In real-world business scenarios, LLM traffic is not deduplicated and naturally includes a massive volume of duplicate benign queries. Without the frozen encoder's exact-match safeguard, the single encoder's aggressive intent clustering would cause these repeated benign requests to be directly judged as malicious, resulting in an absolutely unacceptable false positive rate. Furthermore, in a real-world deployment, an adaptive attacker possessing oracle knowledge of these high-frequency instructions could deliberately exploit this vulnerability, rendering the actual operational degradation even more severe.

\noindent \textbf{The Necessity of Asymmetric Contrastive Learning.}
To confirm that our defense's efficacy stems specifically from the ACL training, we evaluate two critical ablations. First, to verify that our specific asymmetric design is necessary over traditional contrastive learning, we trained a variant using a symmetric contrastive loss where all constituent intents (both monolithic and decomposed slices) are clustered together equally without the anchor-based asymmetric push. As shown by the \textit{Symmetric CL} curve in Figure~\ref{fig:ablation}, this approach results in a severe performance collapse, proving that simply pulling related fragments together destroys the fine-grained semantic boundaries required for accurate detection. Second, to verify that the ACL training itself is the source of the gain, we ablate ACL while preserving the two-encoder architecture. Naively skipping ACL training would force both encoders to share the identical frozen backbone, collapsing the system into a single-encoder pipeline and conflating the contributions of ACL training and architectural redundancy. To avoid this degeneracy, we substitute the ACL-finetuned encoder with a structurally distinct pre-trained model, DeBERTaV2-XLarge~\cite{he2021debertadecodingenhancedbertdisentangled}, so that two heterogeneous encoders coexist without any contrastive fine-tuning. As illustrated by the \textit{w/o ACL} curve in Figure~\ref{fig:ablation}, the resulting Recall-FPR trade-off performs vastly worse than our main TwinGate results. This explicitly demonstrates that even with a near-billion-parameter encoder, raw semantic similarity metrics are fundamentally incapable of bridging the extreme semantic gap between the disjointed fragments of a decompositional jailbreak. Explicit intent-level clustering via ACL is indispensable.

\noindent \textbf{The Role of the Semantic Anchor.}
Finally, we ablate the training data composition by removing the original, complete malicious intent during the ACL fine-tuning phase. In this configuration, the model is trained using only the contrastive relationships among the decomposed fragments themselves. The evaluation yields a Recall-FPR curve (labeled \textit{w/o Raw Intent} in Figure~\ref{fig:ablation}) that is substantially inferior to the primary TwinGate model. This confirms our hypothesis presented in Section~\ref{sec:training_dynamics}: without the complete monolithic intent acting as a dense anchor in the high-dimensional latent space, semantically disparate fragments lack a unifying target to converge upon. Consequently, the resulting clusters suffer from weak boundaries, severely crippling the model's zero-shot generalization capabilities against unseen decompositional jailbreaks.

\noindent \textbf{Generalization to Unseen Decomposers.} To demonstrate TwinGate's zero-shot generalization capabilities against out-of-distribution data, we evaluate its performance when a specific decomposer is entirely excluded from the training phase. Specifically, we removed all decompositional data generated by the Qwen3-30B-A3B model from our training set and evaluated the system on a test set containing attacks exclusively from this unseen decomposer. As illustrated by the \textit{Unseen Decomposer} curve in Figure~\ref{fig:ablation}, TwinGate exhibits only a marginal performance degradation compared to the full model, maintaining a highly competitive Recall-FPR trade-off. This confirms that the ACL framework effectively captures the fundamental semantic trajectories of decompositional jailbreaks, rather than merely overfitting to the idiosyncratic generation patterns of a specific attacker model.

\section{Conclusion}
\label{sec:conclusion}

We introduced TwinGate, a dual-encoder framework that defends against decompositional jailbreaks in high-throughput, untraceable LLM traffic. Where stateless filters fail against temporally disjointed fragments, TwinGate replaces in-context analysis with scalable stateful vector retrieval: ACL bridges the semantic gap by clustering disparate fragments around their shared malicious intent, while a parallel frozen encoder enforces decision inheritance to strictly bound the False Positive Rate on repeated benign traffic.
We additionally curate and release a large-scale benchmark of over 3.62 million requests, spanning 8,600 malicious intents independently decomposed by multiple splitter models, 250k benign intents, and 603k unfragmented samples, providing a reproducible foundation for stress-testing future stateful defenses under strictly causal, deployment-realistic conditions.
Extensive empirical evaluations on this benchmark demonstrate TwinGate's superiority. It achieves a Malicious Intent Recall of over 0.76 while maintaining an exceptionally low FPR of $<0.2\%$, vastly outperforming both stateless 7B-class guardrails and existing stateful monitors. Furthermore, implemented on a hardware-aware, 4-GPU topology utilizing NVLink and in-HBM retrieval, TwinGate sustains a throughput of over 1700 QPS with acceptable $P_{99}$ latency, proving its viability for large-scale production environments. Comprehensive ablations and adaptive attack simulations confirm TwinGate's resilience against both evasion and semantic DoS pollution, establishing it as a practical, scalable defense for production LLM APIs.

\bibliographystyle{ACM-Reference-Format}
\bibliography{sample-base}

\end{document}